\documentclass[preprint,5p,10pt]{elsarticle}
 
\journal{Physics Letters B}

\usepackage{graphicx}
\usepackage{amssymb}

\def\Acknowledgments{\bigskip\bigskip\begin{center}
                     \large\bf Acknowledgments\end{center}}

\def\csuma{Institut f\"ur Theoretische Physik E, RWTH Aachen University,\\
           D-52056 Aachen, Germany}

\def\csumc{Theory Group, Physics Department, CERN,\\
           CH-1211 Geneva 23, Switzerland}

\def\csumd{Physics Department, New York City College of Technology,\\
           300 Jay Street, Brooklyn NY 11201, USA}

\def\eqref#1{(\ref{#1})}
\def\text{\rm }

\begin{document}

\begin{frontmatter}

\title{NLO QED Corrections to Hard-Bremsstrahlung 
         Emission in Bhabha Scattering}

\author{Stefano Actis} 
\address{\csuma}

\author{Pierpaolo Mastrolia}    
\address{\csumc}

\author{Giovanni Ossola}        
\address{\csumd}

\begin{abstract}
    We present a numerical implementation of the one-loop QED corrections 
    to the hard-bremsstrahlung process  $e^- e^+ \to e^-  e^+ \gamma$. These corrections 
    can be included in the Monte Carlo event generators employed for simulating Bhabha 
    scattering events at low-energy high-luminosity electron-positron colliders. The 
    calculation is performed by employing the reduction method developed by Ossola, 
    Papadopoulos and Pittau. Our results are implemented in a modular code for the 
    numerical evaluation of the scattering amplitudes for any given phase-space point. 
    In a similar way, we evaluate also the one-loop QED corrections to $e^- e^+ \to \mu^-  
    \mu^+ \gamma$, and show an interesting application of the method in the presence of 
    two different mass scales in the loops.
\end{abstract}

\end{frontmatter}

\section{Introduction}
Bhabha scattering ($e^- e^+ \to e^- e^+$) is the process employed as a 
luminosity candle at electron-positron colliders. In particular, at 
high-energy colliders such as LEP, the luminosity was measured by 
considering Bhabha scattering events at small scattering angles. At 
high-luminosity meson factories ({\sc DA$\Phi$NE}, {\sc CESR}, etc.)
operating at center-of-mass energies of $1-10$ GeV, instead, the 
luminosity is determined by analyzing large-angle Bhabha scattering 
events. In both cases the Bhabha scattering cross section is large and 
dominated by electromagnetic interactions. Therefore it is  possible to 
employ the techniques of perturbative QED and predict the cross section 
with high accuracy. The smaller theoretical uncertainty which 
affects the resulting cross-section prediction directly translates into 
a more accurate determination of the collider luminosity.

For phenomenological studies and luminosity determinations, fixed-order
calculations of the Bhabha scattering cross section need to be interfaced 
with sophisticated Monte Carlo (MC) generators, which take into account 
realistic experimental cuts and the geometry of the detectors. The current experimental 
analyses rely on programs such as {\sc Babayaga@NLO}~\cite{CarloniCalame:2000pz},  
{\sc MCGPJ}~\cite{Arbuzov:2005pt} and {\sc BHWIDE}~\cite{Jadach:1995nk}, 
which include next-to-leading order (NLO) corrections and effects related 
to multiple photon emission. The three generators mentioned above agree 
within $0.1 \%$ for integrated cross sections and within $1 \%$ for 
distributions as shown in Ref.~\cite{Balossini:2008ht}.

From a fixed-order-calculation perspective, NLO corrections to 
$e^- e^+ \to e^- e^+$ are well under control. In particular, 
one-loop corrections in the full Standard Model were calculated 
long ago~\cite{one-loop}.

NNLO QED corrections to the Bhabha scattering cross section play a 
key role in establishing the accuracy of current MC generators and, 
eventually, in improving it below $0.1 \%$ as it might be required by 
electron-positron colliders of the next generation, such as the  planned 
International Linear Collider.

It is possible to subdivide NNLO corrections into three different sets:  
{\em i}) two-loop corrections to the  process $e^- e^+ \to e^- e^+$; 
{\em ii}) one-loop corrections including a single hard photon in addition  
to the outgoing electron-positron pair;  {\em iii}) tree-level corrections 
with two hard photons or a hard electron-positron pair in addition
to the $e^- e^+$ couple in the final state of Bhabha scattering. 
Obviously, some of the diagrams belonging to the sets {\em i}) and {\em ii}) 
are infrared divergent. However, it is possible in both cases to include a 
class of real soft-photon corrections which can be added to the 
corresponding loop corrections in order to get an infrared-finite result for 
each class.

The calculation of the two-loop corrections belonging to set {\em i}) 
was completed only recently~\cite{Bonciani:2004gi,Penin:2005kf,Becher:2007cu,Actis:2007fs}. Two-photon corrections of class {\em iii}) 
can be computed with any of the existing publicly available  tree-level 
event generators and are safely under control. Radiative 
corrections of class {\em ii}) involve pentagon diagrams with one 
additional hard photon in the final state and are not completely known. 
Partial results are available for small-angle Bhabha scattering~\cite{Arbuzov:1995qd} 
and $s-$channel annihilation  processes at large angles~\cite{Jadach:2001jx}.

The aim of this letter is to present the calculation of the one-loop corrections
belonging to set {\em ii}). Specifically,  we have realized a modular {\sc FORTRAN
95} code which allows for a numerical evaluation at fixed phase-space points of the 
one-loop QED corrections to the  process $e^- e^+ \to e^- e^+ \gamma$ (here $\gamma$ 
denotes a photon with an  energy larger than a given cut-off threshold) retaining a 
finite electron mass. The calculation has been carried out by employing the  
Ossola-Papadopoulos-Pittau (OPP)  method~\cite{opp1,opp2}, based on a reduction of 
the tensor integrals performed at the integrand level~\cite{intlevel}. 

As a by-product, we have computed the one-loop QED corrections to the annihilation 
process $e^- e^+ \to \mu^- \mu^+ \gamma$. This reaction  is an important
background  for the determination of the pion form factor and provides an
independent calibration for a measurement of the hadronic production cross
section.  In addition, it represents an interesting application of the OPP 
reduction method in presence of two mass scales inside loop diagrams. Note that 
a partial implementation of NLO radiative corrections to $e^- e^+ \to \mu^- 
\mu^+ \gamma$ is currently available  in {\sc PHOKHARA}~\cite{Czyz:2004rj} 
and {\sc KK MC}~\cite{Jadach:1999vf} (see Ref.~\cite{Jadach:2006fx} for 
a comparison).

The paper is organized as follows. In Section~\ref{technique} we set our
notation and conventions and briefly describe the computational technique 
which we employed. In Section~\ref{results} we provide numerical results 
for specific benchmark phase-space points.  Section~\ref{end} contains our 
conclusions, including comments on the numerical stability of the results 
and on the computer time required by the calculation.

\section{Calculation of Radiative Corrections}
\label{technique}
\subsection{Conventions and Leading Order Cross Section}
We consider the hard-bremsstrahlung processes
\begin{equation} \label{eq:proc}
  e^-_{k_1} + e^+_{k_2} \to f^-_{k_3} + f^+_{k_4} + \gamma_{k_5},
  \quad  f=e,\mu,
\end{equation}
where $k_1$ and $k_2$ denote the momenta of the colliding electron and
positron, $k_3$ and $k_4$ stand for the momenta of the outgoing fermion 
and anti-fermion, and $k_5$ is the momentum of the outgoing hard photon. 
The particle momenta obey the mass-shell conditions $k_1^2=k_2^2=m_e^2$, 
$k_3^2=k_4^2=m_f^2$ and $k_5^2=0$. The kinematics is described by five 
independent invariants, which can be  chosen among the six quantities
\begin{eqnarray} \label{eq:Man}
  s &=& (k_1+k_2)^2 = 4 E^2 \ , \nonumber \\ 
  s'&=& (k_3+k_4)^2 \ , \\ 
  t_{ij}&=& (k_i-k_j)^2 \ , \qquad (i=1,2; \ j=3,4). \nonumber
\end{eqnarray}
Here $E$ indicates the beam energy in the center-of-mass frame.

\begin{figure}[t]
\begin{center}
  \includegraphics[scale=0.35]{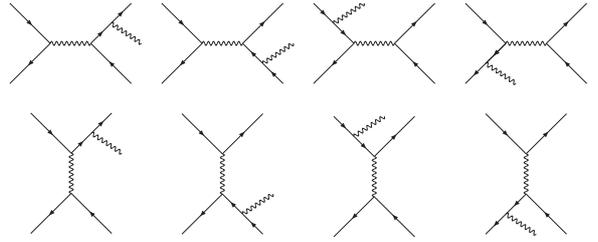}
  \caption[]{QED tree-level diagrams for $e^- e^+ \to e^- e^+ \gamma$ 
             (full set) and $e^- e^+ \to \mu^- \mu^+ \gamma$ (first
             line). Here and in the following, particles on the left 
             side of the diagrams are considered incoming, particles 
             on the right side of the diagrams are outgoing.}
  \label{fig:tree}
\end{center}
\end{figure}

The unpolarized leading order (LO) cross section can be written as
\begin{equation} \label{eq:LO}
  d\sigma_{\rm LO}=
  \frac{1}{2 \sqrt{s\, (s-4 m_e^2)}}\, d R_3\,
  \frac{1}{4} \sum_{\rm spins} | {\cal M}_{\rm tree} |^2,
\end{equation}
where ${\cal M}_{\rm tree}$ represents the tree-level amplitude. 
In the context of pure QED, the latter originates from the 
Feynman diagrams shown in Fig.~\ref{fig:tree}. The $n$-particle 
phase space is defined as
\begin{equation} \label{eq:PS}
\hspace*{-0.7cm}
 d R_n= \left( \prod_{i=3}^{n+2} \frac{d^3 k_i}{(2\pi)^3 2 E_i} \right)
  (2\pi)^4 \delta^{(4)}\left(k_1 \!+\! k_2 - \sum_{j=3}^{n+2} k_j\right)\!.
\end{equation}

The LO cross section for real-photon emission of Eq.~\eqref{eq:proc} 
was computed long ago in Refs.~\cite{Swanson:1966ks}, and in the ultra-relativistic 
limit it can be written in a compact way~\cite{Berends:1981rb}. We report its 
expression because it can be used for checks, although our NLO results, presented in the
following sections, have been computed without any approximations, by retaining 
the complete dependence on the kinematical variables and keeping finite electron and muon masses.
The ultra-relativistic limit of the LO results \cite{Berends:1981rb} are conveniently expressed in 
terms  of the kinematic invariants of Eq.~\eqref{eq:Man} and of the 
variables $k_{i5}= k_i \cdot k_5$, with $ i=1,\ldots,4$,
\begin{eqnarray}
&& \hspace*{-1.2cm}
 \frac{1}{4} \sum_{\rm spins} | {\cal M}_{\rm tree} |^2 = 
 \left( 4 \, \pi \, \alpha \right)^3 
  \, {\cal T} \, 
 \Big( \frac{s}{k_{15}k_{25}}   +
        \frac{s'}{k_{35}k_{45}} + \nonumber \\
       && 
      - \frac{t_{13}}{k_{15}k_{35}}  -
        \frac{t_{24}}{k_{25}k_{45}}  +
        \frac{t_{14}}{k_{15}k_{45}}  +
        \frac{t_{23}}{k_{25}k_{35}}    \Big),
\end{eqnarray}
where the function ${\cal T}$ depends on the final-state fermions. 
For the process $e^- e^+ \to e^- e^+ \gamma$ one finds 
\begin{eqnarray}
&& \hspace*{-1.2cm}
  {\cal T}= \frac{1}{s s' t_{13} t_{24}} \Big[ 
  s s' (s^2+s'^2) + 
  t_{13} t_{24} (t_{13}^2 + t_{24}^2 ) + \nonumber \\ && \qquad \qquad + 
  t_{14} t_{23} (t_{14}^2 + t_{23}^2 )
  \Big],
\end{eqnarray}
while for the case $e^- e^+ \to \mu^- \mu^+ \gamma$ the function ${\cal T}$ is
\begin{equation}
  {\cal T}= \frac{1}{s s'} \left( 
  t_{13}^2 + t_{14}^2  +t_{23}^2 + t_{24}^2 \right).
\end{equation}
\subsection{Virtual Corrections}
\label{virtual}
\begin{figure}[ht]
\begin{center}
   \includegraphics[scale=0.35]{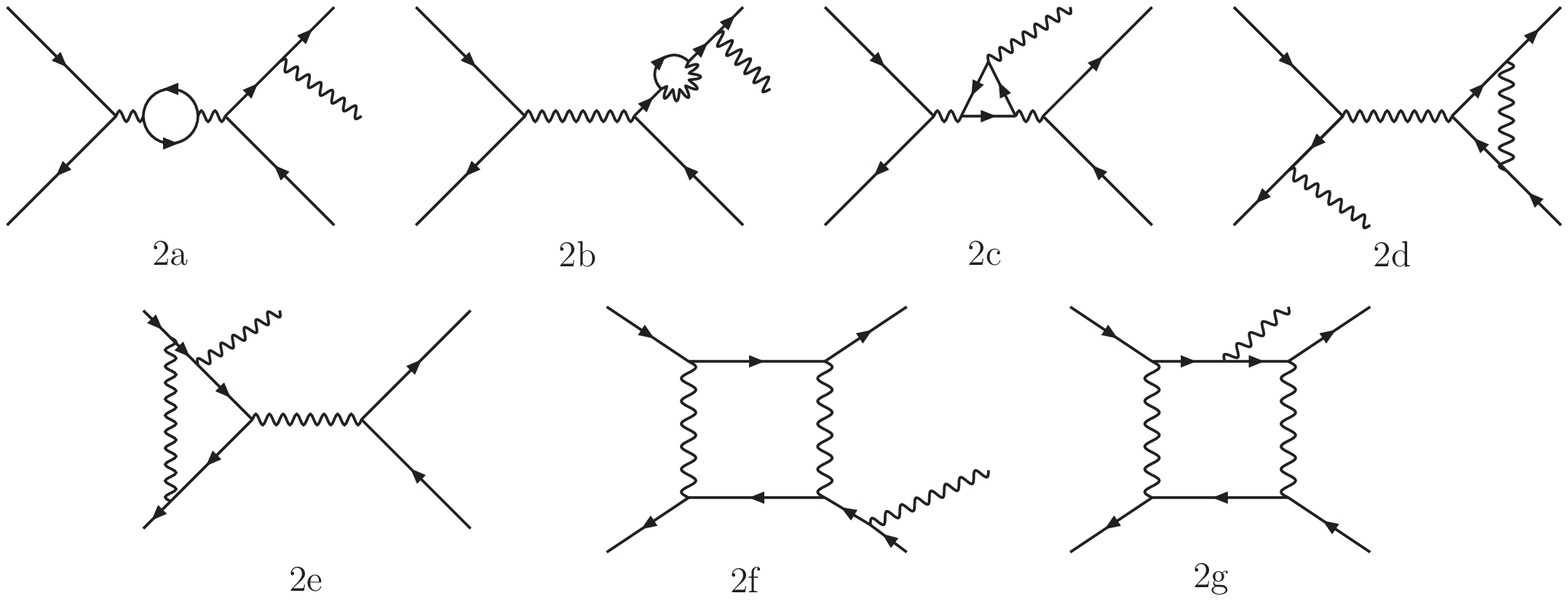}
   \caption[]{Representative one-loop diagrams for $e^- e^+ \to 
              e^- e^+ \gamma$.}
   \label{fig:1loop}
\end{center}
\end{figure}

The one-loop corrections to the two scattering processes studied
in this work involve a relatively limited number of Feynman diagrams:
the package {\sc QGRAF}~\cite{qgraf} generates 38 one-loop diagrams 
for the process $e^- e^+  \to \mu^- \mu^+ \gamma$ and 76 diagrams for 
the process $e^- e^+ \to e^- e^+ \gamma$. Representative graphs are 
shown in Fig.~\ref{fig:1loop}; note that due to Furry's theorem, 
diagrams of class 2c cancel in the sum.

As can be seen from Fig.~\ref{fig:1loop}, the calculation of the 
complete one-loop corrections requires the evaluation of pentagon 
diagrams of class 2g, the most challenging and time consuming part of this 
calculation.

It is well known that any one-loop amplitude can be written as the sum of
a linear combination of scalar box, triangle, bubble and tadpole integrals, 
whose analytic expressions were obtained  in the seminal paper by 't Hooft 
and Veltman~\cite{thv}, and a so-called rational part. The rational term and 
the coefficients multiplying the one-loop integrals are ratios of polynomials 
in the masses and the kinematical invariants. These coefficients have been 
traditionally determined by means of the Passarino-Veltman (PV) reduction 
method~\cite{PV}. However, for processes with more than four external 
legs, a straightforward application of the PV reduction can generate very 
large expressions involving inverse Gram determinants and lead to numerical 
instabilities. 

In recent years, a significant amount of work has been devoted to the 
evaluation of one-loop amplitudes for processes with five or more 
external legs. Besides improvements on standard techniques, where a 
tensor reduction is explicitly performed, new unitarity-based numerical 
and analytical developments have emerged (for recent reviews see 
Refs.~\cite{Bern:2008ef} where an exhaustive list of  references can be 
found). Both directions have been successfully pursued and have led to 
impressive results. In particular, in the last few months, very challenging 
calculations involving four particles in the final state, relevant for LHC phenomenology, 
have been completed~\cite{tt1,w3j}.

In the present letter we describe the evaluation of the one-loop corrections 
to the processes $e^- e^+  \to e^- e^+ \gamma$ and $e^- e^+ \to \mu^- 
\mu^+ \gamma$, performed by means of the OPP reduction method~\cite{opp1,opp2}. 
The one-loop amplitudes, generated through {\sc QGRAF}, have been processed 
with {\sc FORM}~\cite{form} routines in order to produce a {\sc FORTRAN 95} output. 
The latter has  then been used as an input for two computer implementations of the OPP 
technique which numerically evaluate the coefficients of the basis integrals
and the rational part for each phase-space point~\cite{cuttools, red}. 
Both codes have been interfaced with publicly available software which allows
for the evaluation of the basis integrals~\cite{qcdloop,olo}.

The one-loop  corrections to the cross section can be written as
\begin{equation}
\hspace*{-0.8cm}
  d \sigma^{\rm V}_{\rm NLO}=
  \frac{d R_3}{2\, \sqrt{s (s-4 m_e^2)}}\, \,
  \frac{1}{4} \sum_{\rm spins} \, 2\,  \text{Re} \, 
  \left( {\cal M}_{\rm 1-loop} {\cal M}^{\star}_{\rm tree} \right),
\end{equation}
where
the interference of the 
one-loop amplitude ${\cal M}_{\rm 1-loop}$ with the complex-conjugate 
tree-level amplitude ${\cal M}^{\star}_{\rm tree}$ is evaluated in 
dimensional regularization and subdivided into four contributions,
according to the OPP-reduction algorithm,
\begin{equation} \label{eq:deco}
\hspace*{-0.8cm}
  \frac{1}{4} \sum_{\rm spins} \, 2 \, \text{Re} \, 
  \left( {\cal M}_{\rm 1-loop} {\cal M}^{\star}_{\rm tree} \right) =
  {\cal CC}_4 + {\cal R}_1 + {\cal R}_2 + {\cal UV}_{ct} \ .
\end{equation}
Here ${\cal CC}_4$ denotes the cut-constructible four-dimensional
part of the result, written as a linear combinations of scalar
boxes, triangles, bubbles and tadpoles, ${\cal R}_1+{\cal R}_2$ 
stands for the so-called rational part and ${\cal UV}_{ct}$ summarizes 
all contributions induced by the ultraviolet counterterms. As a
technical remark, we stress that throughout our computation we have 
used strictly four-dimensional external momenta.

The rational part has been written through the sum of the components 
${\cal R}_1$ and ${\cal R}_2$, as explained in Ref.~\cite{Ossola:2008xq}. 
The integrand corresponding to a generic $m-$point one-loop Feynman 
diagram can be written as
\begin{equation} \label{lala}
\hspace*{-0.5cm}
  \bar{A}(\bar{q}) = \frac{\bar{N}(\bar{q})}
  {\bar{D}_0 \cdots \bar{D}_{m-1}}, \qquad
  \bar{D}_i= \left(\bar{q}+p_i\right)^2-m_i^2 \ ,
\end{equation}
where a bar denotes objects living in $d=4-2 \epsilon$ 
space-time dimensions, $q$ is the loop momentum, $p_i$ are linear 
combinations of the external four-dimensional momenta and $m_i$ stand 
for the masses of the internal legs. The numerator $\bar{N}(\bar{q})$ 
can be split into a four-dimensional and an $\epsilon$-dimensional part, 
denoted in the following by a tilde, $\bar{N}(\bar{q})=N_1(q)+\tilde{N}_2(\bar{q})$. 
The four-dimensional numerator $N_1(q)$ is then expanded in terms of 
four-dimensional denominators $D_{i}$ \cite{opp1}: the mismatch in the 
dimensionality between the latter and the $d-$dimensional inverse propagators 
$\bar{D}_i$ of Eq.~\eqref{lala} generates the ${\cal R}_1$ term in 
Eq.~\eqref{eq:deco}. The ${\cal R}_2$ component, instead, stems from the 
overlap between  $\tilde{N}_2(\bar{q})$ and the ultraviolet poles of 
one-loop integrals, and can be evaluated in QED by employing the {\it ad-hoc} 
counterterm-like Feynman rules presented in Ref.~\cite{Ossola:2008xq}. Note 
that the overlap between the $\epsilon$-dimensional numerator 
$\tilde{N}_2(\bar{q})$ and the infrared poles of one-loop integrals can be 
safely neglected from the very beginning as proven in Appendix~A of 
Ref.~\cite{Bredenstein:2008zb}.

We carried out the ultraviolet renormalization in the on-mass-shell
scheme: the renormalized charge is chosen to be equal to the value
of the electromagnetic coupling, as measured in Thomson scattering, 
at all orders in perturbation theory; the squared fermion masses are 
identified with the real parts of the poles of the Dyson-resummed 
propagators; field-renormalization constants cancel by definition 
external wave-function corrections.

\begin{figure}[ht]
\begin{center}
\includegraphics[scale=0.3]{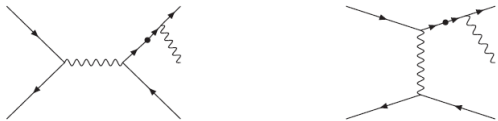}
\caption[]{Representative mass-counterterm diagrams for 
           $e^- e^+ \to e^- e^+ \gamma$. Black dots stand 
           for mass-counterterm insertions, necessary for performing
           renormalization and computing the rational term ${\cal R}_2$
           in the massive case.}
\label{fig:ctmass}
\end{center}
\end{figure}

Throughout our computation we have retained the full dependence on 
the fermion masses, introducing in particular the appropriate 
ultraviolet and ${\cal R}_2$-type mass counterterm diagrams depicted 
in Fig.~\ref{fig:ctmass}. 
We observe that, in the limit where terms suppressed by positive 
powers of the electron or muon mass are neglected, the contributions 
induced by ultraviolet counterterms and ${\cal R}_2$-like rational 
terms can be embedded in the computation by means of simple factors multiplying 
the tree-level amplitude. They represent important 
elements for performing checks, as they can be accounted for simply by
introducing counterterms at the diagrammatic level. Moreover,
they can be employed for speeding up the numerical evaluation
of the amplitudes in all cases where terms proportional to the 
fermion masses can be safely neglected.

By using the U(1) Ward identity in order to eliminate the charge 
counterterm, we find that the ultraviolet counterterm ${\cal UV}_{ct}$ 
in Eq.~(\ref{eq:deco}) originates from
\begin{equation} \label{eq:qui1}
  {\cal M}_{\rm 1-loop}\left[{\cal UV}_{ct}\right]= {\cal M}_{\rm tree}
  \frac{\alpha}{4 \pi} 
  \left( \, 2\, \delta {\cal Z}_\psi - \delta {\cal Z}_A \, \right),
\end{equation}
where field counterterms are defined through $A_\mu= {\cal Z}_A^{1\slash 2} 
A_\mu^{\rm R}$, $\psi= {\cal Z}_\psi^{1\slash 2} \psi^{\rm R}$,
${\cal Z}_i = 1 + \alpha\slash (4\pi) \delta {\cal Z}_i$, with $i=A,\psi$, 
and the index $\rm R$ has been introduced for denoting renormalized quantities.
Explicitly, we have the well-known expressions
\begin{eqnarray} \label{eq:q1}
\hspace*{-0.7cm}
\delta {\cal Z}_\psi\!&=&\! Q_f^2 \left\{ -3 \left[ \frac{1}{\epsilon} - \gamma - 
{\rm{\ln}} \pi -  \, {\rm{\ln}} \left(\frac{m_f^2}{\mu^2}\right) 
\right]- 4\right\},\nonumber \\
\hspace*{-0.8cm} && \\
\hspace*{-0.8cm}
\delta {\cal Z}_A\!&=&\! \frac{4}{3} \sum_f N_f Q_f^2 
\left[ -\frac{1}{\epsilon} + \gamma + 
{\rm{\ln}} \pi  +  {\rm{\ln}} \left(\frac{m_f^2}{\mu^2}\right) \right],
\nonumber
\end{eqnarray}
where the index $f$ runs over the fermions, $N_f$ is the color factor ($N_f=1$ for
leptons and $N_f=3$ for quarks), $Q_f$ is the quantum number associated
with the electric charge, $m_f$ is the fermion mass, $\gamma$ is the 
Euler-Mascheroni constant and $\mu$ is the 't Hooft mass. 
Although QED is not simultaneously free from ultraviolet and
infrared divergencies for an arbitrary number of space-time
dimensions, we have followed the common practice of using the same 
dimensional regulator for isolating both ultraviolet and infrared poles.

Similarly, within the same approximation, the ${\cal R}_2$ component in 
Eq.~(\ref{eq:deco}) arises from the interference of ${\cal M}_{\rm 1-loop}\left[ {\cal R}_2 \right]$ 
with the tree-level amplitude, where the former is defined as,
\begin{equation} \label{eq:qua2}
  {\cal M}_{\rm 1-loop}\left[ {\cal R}_2 \right]= {\cal M}_{\rm tree}
  \frac{\alpha}{4 \pi} 
  \Bigl( \, -5 + \frac{2}{3} \sum_f N_f Q_f^2  \, \Bigr) .
\end{equation}
In both Eq.~\eqref{eq:qui1} and 
Eq.~\eqref{eq:qua2}, vacuum polarization insertions can be easily removed by 
switching off all terms proportional to the sum over the fermions.

We observe that the common structure of those terms, even in the general kinematic case we worked out, 
suggests that the contributions to the rational term 
${\cal R}_2$ can be fully combined with ultraviolet counterterms, thus 
achieving a significant optimization in this part of the calculation.

The cut-constructible term and the ${\cal R}_1$ term in 
Eq.~(\ref{eq:deco}) have been obtained by direct application of the OPP 
method. Here we describe the two implementations of this part of
the calculation. 

The first version employs the routines of the publicly available package
{\sc CutTools}~\cite{cuttools} for the numerical determination of the 
coefficients which multiply the basis integrals. These routines have been 
combined with {\sc QCDLoop}~\cite{qcdloop}, used to evaluate the needed 
scalar integrals~\footnote{We have used the version 1.0 of {\sc CutTools}. 
During the completion of this work, version 1.1 has become available at 
\texttt{http://www.ugr.es/$\sim$pittau/CutTools/}, with a built-in interface 
to {\sc QCDLoop}.}.
The basis of loop integrals employed by {\sc CutTools} 
involves also rank-one and rank-two two-point functions in order to improve 
the numerical stability respect to the case where the basis contains only 
scalar integrals, as described in Section~3 of Ref.~\cite{opp2}. The additional 
tensor integrals  have been included coding the relations provided in
Ref.~\cite{Denner:2005nn}.

The second version of the calculation is used for cross-checking our results
and it makes use of an independent {\sc FORTRAN 95} code~\cite{red} for the reduction 
of the tensor integrals, which includes an optional optimization of the 
OPP technique based on the Discrete Fourier Transform~\cite{Mastrolia:2008jb}. 
The basis integrals are evaluated using the code {\sc OneLOop}~\cite{olo}, written by 
A.~van Hameren. 

It is interesting to observe that the rational part ${\cal R}_1$ is 
computed by using different strategies in the two versions of the calculation: 
{\sc CutTools} employs the mass-shift procedure introduced  in Ref.~\cite{opp1}, 
while the second approach uses the counterterm-based method described in 
Ref.~\cite{opp2}.

The results obtained with the two independent implementations for the 
cut-constructible term and the component ${\cal R}_1$ are in very good agreement. 

A final comment concerns the vacuum polarization insertions  (such as diagram 2a
in  Fig.~\ref{fig:1loop}). Although these types of diagrams are implemented in both
codes, they are naturally incorporated in any QED computation by simply running
the fine-structure constant $\alpha$ to the appropriate scale. Furthermore, the
introduction of a running fine-structure constant allows for a straightforward
inclusion of hadronic contributions using dispersion relations and the optical
theorem. Since this class of corrections is  well-known, it will be neglected in
the rest of the paper.
\subsection{Real Soft-Photon Corrections} 
\label{soft}
In order to check our calculation, it is useful to control the 
agreement between the values of the coefficients of the residual 
infrared poles for the virtual one-loop corrections and those of 
the analytic expressions of the corresponding poles arising from 
real soft-photon emission diagrams. In fact, the sum of the one-loop corrections to $e^- e^+ 
\to e^- e^+ \gamma$ ($e^- e^+ \to \mu^- \mu^+ \gamma$) and the 
real-emission diagrams for $e^- e^+ \to e^- e^+ \gamma \, 
\gamma_{\mbox{{\tiny soft}}}$ ($e^- e^+ \to \mu^- \mu^+ \gamma \, 
\gamma_{\mbox{{\tiny soft}}}$) is infrared finite after integrating over 
the soft-photon phase space up to a given upper cut-off on the 
energy of the undetected photon $\gamma_{\mbox{{\tiny soft}}}$.

The contribution of real soft-emission diagrams factorizes in the product
of the LO cross section of Eq.~\eqref{eq:LO} and an infrared-divergent 
coefficient which can be extracted from Ref.~\cite{BP},
\begin{equation}
  d \sigma^{\rm R}_{\rm NLO}=  \frac{\alpha}{\pi} \,
  d \sigma_{\rm LO} \, \left( \frac{1}{\epsilon} \sum_{i,j=1}^4 J_{ij}
  + \Delta J_{ij} \right),
\end{equation}
where the functions $J_{ij}$ can be written in the compact form
\begin{equation} \label{JJ}
  J_{ij} = \frac{\varepsilon_i \varepsilon_j}{4 \beta_{ij}} \mbox{ln}
           \left(\frac{1+\beta_{ij}}{1-\beta_{ij}}\right), 
\end{equation}
with
\begin{equation}
    \beta_{ij}= \sqrt{1-\frac{m_i^2 m_j^2}{(k_i \cdot k_j )^2}},
\end{equation}
and $\Delta J_{ij}$ being a finite remainder which is not considered in the
present work. Here we have introduced $\varepsilon_i=1$ for $i=1,4$ and 
$\varepsilon_i=-1$ for $i=2,3$. In the case $i=j$, obtained by taking the 
limit $\beta_{ij} \to 0$ in Eq.~(\ref{JJ}), one finds $J_{ii}= 1 \slash 2$.
\section{Numerical Results}
\label{results}
In this section we show the numerical results for the squared LO amplitude,
summed and averaged over the spins, and the associated one-loop virtual 
corrections (excluding trivial vacuum polarization insertions) for both 
processes at fixed phase-space points. Although the main result of our
work is represented by a computer code which can eventually be
interfaced with MC generators, we find it useful to show explicitly
some numerical results in order to allow for detailed comparisons.

We define
\begin{equation} \label{defa}
  {\cal I}_{\rm LO} = \frac{1}{4} \sum_{\rm spins} | {\cal M}_{\rm tree} |^2 \ ,
\end{equation}
\begin{equation}\label{defb}
\left( e^{\gamma} \pi \right)^{-\epsilon} \, {\cal I}^{\rm V}_{\rm NLO}
 =  \frac{1}{4} \sum_{\rm spins} \, 2 \, \text{Re} \,
  \left( {\cal M}_{\rm 1-loop} {\cal M}^{\star}_{\rm tree} \right) \ ,
  \end{equation}
where  $\gamma= 0.5772156 \cdots$ and the arbitrary 
't Hooft mass unit has been set for definiteness to the value $\mu=1$ GeV. 
Following Eq.~(\ref{eq:deco}), we further define
\begin{equation} \label{asplit}
{\cal I}^{\rm V}_{\rm NLO} = 
{\cal I}^{\rm V}_{\rm NLO}({\cal CC}_4 +{\cal R}_1 + {\cal R}_2)+
{\cal I}^{\rm V}_{\rm NLO}({\cal UV}_{ct}),
\end{equation}
isolating the contribution of the ultraviolet counterterms from the sum
of the cut-constructible and rational parts. 

In the following we will show 
also the value of the residue the infrared pole for real soft-photon emission,
introducing
\begin{equation}
{\cal I}^{\rm R}_{\rm NLO} = \frac{\alpha}{\pi} {\cal I}_{\rm LO}
\frac{1}{\epsilon} \sum_{i,j=1}^4 J_{ij},
\end{equation}
with $J_{ij}$ defined in Eq.~\eqref{JJ}.

All results have been obtained using the input data suggested by the Particle 
Data Book~\cite{PDB}.
\subsection{Results for $e^- e^+ \to \mu^-  \mu^+ \gamma $} \label{muon-sec}
We start considering the process $e^- e^+ \to \mu^-  \mu^+ \gamma $.
We set $\sqrt{s}= 10$ GeV and show results for the following phase-space 
point:
\scriptsize
\begin{eqnarray}
\hspace*{-0.5cm}
  k_1 \!\!\!&=&\!\!\! (5,0, 4.999999973888011,0),\nonumber\\
\hspace*{-0.5cm}
  k_2 \!\!\!&=&\!\!\!(5,0,-4.999999973888011,0),\nonumber\\
\hspace*{-0.5cm}
  k_3 \!\!\!&=&\!\!\! (0.5738577925797953,
           0.4550096995790255, \nonumber \\ && \quad
          -0.3315217520850587,
           3.476459076282246 \cdot 10^{-2}), \nonumber\\
\hspace*{-0.5cm}
 k_4 \!\!\!&=&\!\!\! (4.947897617027284,
         -1.988746555731907, \nonumber \\ && \quad
          4.244475358996690, 
          -1.581089245425936),\nonumber\\
\hspace*{-0.5cm}
 k_5 \!\!\!&=&\!\!\! (4.478244590392921,
          1.533736856152882, \nonumber \\ && \quad
         -3.912953606911631,
          1.546324654663113),  
\end{eqnarray}
\normalsize
where $k_i=(k_i^0,k_i^1,k_i^2,k_i^3)$ and all quantities are expressed in GeV. 
After splitting the one-loop corrections according to Eq.~(\ref{asplit}), we obtain:
\scriptsize
\begin{eqnarray}
\label{lalat}
\hspace*{-0.5cm}
 {\cal I}_{\rm LO} \!\!\!&=&\!\!\! 5.013964825924999 \cdot 10^{-3} ,\nonumber\\
\hspace*{-0.5cm}
 {\cal I}^{\rm V}_{\rm NLO} ({\cal CC}_4 \!+\! {\cal R}_1 \!+\!{\cal R}_2) \!\!\!&=&\!\!\! 
   \frac{1}{\epsilon} \,  \, 0.3666265876159401 \cdot 10^{-3} + \nonumber \\ && \quad
  + 1.944055391172138 \cdot 10^{-3},\nonumber\\
\hspace*{-0.5cm}
 {\cal I}^{\rm V}_{\rm NLO} ({\cal UV}_{ct}) \!\!\!&=&\!\!\!
  - \, \frac{1}{\epsilon} \,\, 0.03481919776738505 \cdot 10^{-3} + \nonumber \\ && \quad
  - 0.3892249340437660 \cdot 10^{-3} ,\nonumber\\
\hspace*{-0.5cm}
 {\cal I}^{\rm V}_{\rm NLO}\!\!\!&=&\!\!\! 
   \frac{1}{\epsilon} \,\, 0.3318073898485551 \cdot 10^{-3} + \nonumber \\ && \quad
  + 1.554830457128372 \cdot 10^{-3} ,\nonumber\\
\hspace*{-0.5cm}
  {\cal I}^{\rm R}_{\rm NLO}\!\!\!&=&\!\!\! -\, \frac{1}{\epsilon} \,\, 0.3318073896945480 
  \cdot 10^{-3}.
\end{eqnarray}
\normalsize
All results are expressed in $\text{GeV}^{-2}$ and they have been obtained by working in double 
precision. 

In order to check the stability of our results, we have performed
the so-called $N=N$ test, monitoring the agreement between 
the numerical values of the numerator function $\bar{N}(\bar{q})$ of Eq.~\eqref{lala} 
before and after the decomposition in terms of inverse propagators~\cite{cuttools}. 
When the numerical agreement in the comparison does not reach a given limit
set by the user, 
the code automatically triggers the use of the more time-consuming multi-precision 
routines~\cite{Bailey}. We observe more than 9 digits of agreement between
the results obtained in double precision, requiring a $10^{-5}$ relative
precision for the $N=N$ test, and those we got after forcing multi-precision 
in the reduction program for reaching a $10^{-15}$ relative precision.

A second test on the precision of our calculation concerns the cancellation of the 
poles. After renormalization, the residual pole for ${\cal I}^{\rm V}_{\rm NLO}$ is of pure 
infrared origin and it matches the infrared pole in ${\cal I}^{\rm R}_{\rm NLO}$
with an agreement of 9 digits.

Finally, as stressed in Section~\ref{virtual}, we have performed the calculation of 
all contributions by means of two independent codes.

As a further test of the stability of our results, we study the process $e^- e^+ \to \mu^-  \mu^+ \gamma $
for configurations in which the muon, or the antimuon, is (almost) parallel to the emitted photon. 
We set $\sqrt{s}= 1$ GeV and fix the momenta of the colliding leptons and the outgoing photon to be:
\scriptsize
\begin{eqnarray} \label{plot:momenta}
\hspace*{-0.5cm}
k_1 \!\!\!&=&\!\!\! (0.5, 0, 0, 4999997388800458),\nonumber\\
\hspace*{-0.5cm}
k_2 \!\!\!&=&\!\!\! (0.5, 0, 0, -0.4999997388800458),\nonumber\\
\hspace*{-0.5cm}
k_5 \!\!\!&=&\!\!\! (0.4, 0.1647604992975971, \nonumber \\ && \quad
                      -0.3568677260727233, 0.07415796625373612),
\end{eqnarray}
\normalsize
with all momenta given in GeV. 

\begin{figure}[h]
\begin{center}
  \includegraphics[scale=0.8]{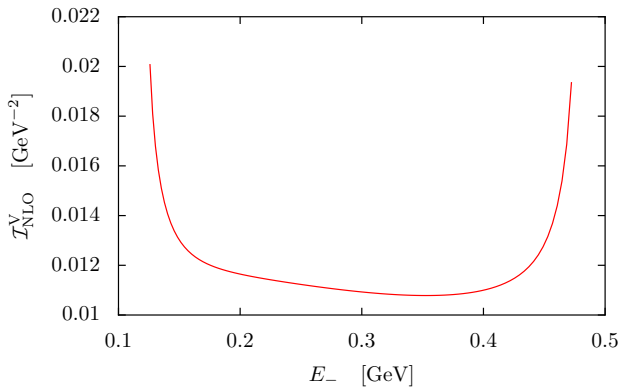}
  \caption[]{Virtual corrections for  $e^- e^+ \to \mu^-  \mu^+ \gamma $
             for $\sqrt{s}= 1$ GeV as a function of the energy of the outgoing muon.
             The momenta of the colliding leptons and the outgoing photon are defined in Eq.~\eqref{plot:momenta}.}
  \label{collinear.eps}
\end{center}
\end{figure}

Then, we calculate the finite part of
the virtual corrections ${\cal I}^{\rm V}_{\rm NLO}$ of Eq.~\eqref{defb},
as a function of the energy of the outgoing muon $E_-$: the results are illustrated in Fig.~\ref{collinear.eps}.
On the left side of the plot, for low values of $E_-$, we are in a configuration in which the 
muon is almost parallel to the photon, while the antimuon goes in the opposite direction;
we then progressively rotate the muon, keeping the four-momentum of the photon fixed, until the muon
becomes anti-parallel to the photon, while the antimuon becomes almost parallel to it.
While we observe the appearance of collinear divergencies, the virtual corrections ${\cal I}^{\rm V}_{\rm NLO}$ follow a smooth curve, thus suggesting a good stability of our result even in this particular kinematic configuration.

\subsection{Results for $e^- e^+ \to e^-  e^+ \gamma $}
Here we consider the process $e^- e^+ \to e^-  e^+ \gamma $. We
set $\sqrt{s}= 1$ GeV and present numerical results for the following 
phase-space point:
\scriptsize
\begin{eqnarray}
\hspace*{-0.5cm}
k_1 \!\!\!&=&\!\!\! (0.5,0, 0.4999997388800458,0),\nonumber\\
\hspace*{-0.5cm}
k_2 \!\!\!&=&\!\!\! (0.5,0,-0.4999997388800458,0),\nonumber\\
\hspace*{-0.5cm}
k_3 \!\!\!&=&\!\!\! (0.1780937847558600,
        -0.1279164180985903, \nonumber\\ && \quad
         5.006809884093004 \cdot 10^{-2},
        -0.1133477415216646 ),
 \nonumber\\
\hspace*{-0.5cm}
k_4 \!\!\!&=&\!\!\! (0.3563944406457374 ,
        -2.860530642319879 \cdot 10^{-2} , \nonumber\\ && \quad
        -0.1832142729949070  ,
        -0.3043534176228102 ),\nonumber\\
\hspace*{-0.5cm}
k_5 \!\!\!&=&\!\!\! (0.4655117745984024,
         0.1565217245217891, \nonumber\\ && \quad
         0.1331461741539769,
         0.4177011591444748),
\end{eqnarray}
\normalsize
with all quantities expressed in GeV.  We obtain
\scriptsize
\begin{eqnarray}
 {\cal I}_{\rm LO} \!\!\!&=&\!\!\! 0.7586101468103619 ,\nonumber\\
 {\cal I}^{\rm V}_{\rm NLO} ({\cal CC}_4 \!+\! {\cal R}_1 \!+\! {\cal R}_2) \!\!\!&=&\!\!\! 
\frac{1}{\epsilon} \,\, 
        0.04745064270035045  + \nonumber \\ && \quad
       + 0.5005828268263969,\nonumber\\
 {\cal I}^{\rm V}_{\rm NLO} ({\cal UV}_{ct}) \!\!\!&=&\!\!\! 
  - \, \frac{1}{\epsilon} \,\, 0.005286348050945757 + \nonumber \\ && \quad
                       - 0.08718044078580632  ,\nonumber\\
 {\cal I}^{\rm V}_{\rm NLO}\!\!\!&=&\!\!\!
           \frac{1}{\epsilon} \,\, 0.04216429464940469 + \nonumber \\ && \quad
                       + 0.4134023860405905   ,\nonumber\\
  {\cal I}^{\rm R}_{\rm NLO}\!\!\!&=&\!\!\! - \, \frac{1}{\epsilon} \, \, 0.04216429464958627,
\end{eqnarray}
\normalsize
where all results are expressed in $\text{GeV}^{-2}$. 

All the numbers have been obtained by working in double precision and requiring a $10^{-5}$ 
relative precision for the $N=N$ test. Also for this process, we have tested the 
precision of our calculation by forcing multi-precision in the reduction program, by 
checking the complete cancellation of ultraviolet and infrared poles and by
comparing the two independent implementations.

\begin{figure}[ht]
\begin{center}
  \includegraphics[scale=0.8]{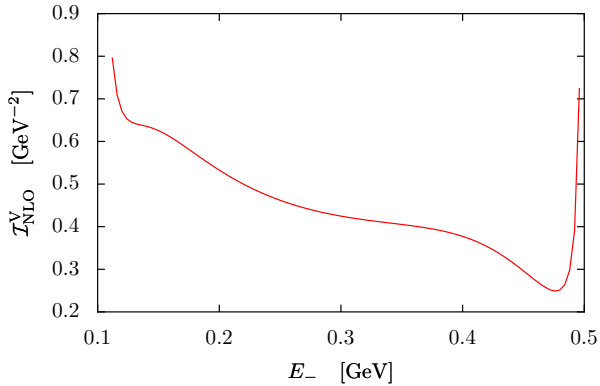}
  \caption[]{Virtual corrections for  $e^- e^+ \to e^-  e^+ \gamma $
             for $\sqrt{s}= 1$ GeV as a function of the energy of the outgoing electron.
             The momenta of the colliding leptons and the outgoing photon are 
             defined in Eq.~\eqref{plot:momenta}}
  \label{collinearbh}
\end{center}
\end{figure}

Following the same idea of Fig.~\ref{collinear.eps},
we plot in Fig.~\ref{collinearbh} the finite part of
the virtual corrections, ${\cal I}^{\rm V}_{\rm NLO}$ of Eq.~\eqref{defb},
as a function of the energy of the outgoing electron $E_-$, ranging
between two configurations where the electron
momentum is (almost) parallel or anti-parallel to the photon momentum
and collinear divergencies show up.
\section{Conclusions} \label{end}
We have applied the OPP method to evaluate the complete NLO virtual QED 
corrections to the hard bremsstrahlung processes $e^- e^+ \to e^-  e^+  
\gamma$ and $e^- e^+ \to \mu^-  \mu^+ \gamma$, both relevant for determining 
the luminosity at low-energy electron-positron colliders.

It is interesting to note that in the case of muon-pair production in association 
with a hard photon the calculation has been performed retaining both the 
electron and muon masses. This shows the capability of the method in dealing 
with several scales inside the loop diagrams.

The numerical stability of the results has been tested by means of
three different procedures: the check on the quality of the reconstructed
numerators given by the $N=N$ test; the comparison of infrared divergencies 
arising from virtual and real corrections, which cancel as they should; the 
agreement of the results obtained using two independent codes. In particular, 
for LO calculations, we find an agreement of at least 12 digits; for NLO results, 
we have 9 digits of accuracy.

The main result of our calculation is the implementation of hard-bremsstrahlung
emission corrections, both for Bhabha scattering and muon-pair production, 
into a {\sc FORTRAN 95} code which employs the publicly available package {\sc CutTools} 
for the extraction of the coefficients of the scalar integrals and the 
package {\sc QCDLoop} for evaluating the needed scalar integrals. 

The typical order of magnitude of the CPU time of the {\sc FORTRAN 95} 
code we have developed is ${\cal O}(10^{-1})$ seconds for each phase-space 
point on a standard desktop machine. The computational speed,
together with the modular approach we have followed in computing the
radiative corrections, allows for an implementation of the results in the
existing MC programs.

Once supplemented with the finite part for real-emission diagrams, the results we have 
derived allow for a phenomenological study of the two hard-bremsstrahlung
processes $e^- e^+ \to e^-  e^+ \gamma$ and $e^- e^+ \to \mu^-  \mu^+ \gamma$,
including the calculation of the cross sections and the relevant distributions.
In addition, they can be compared with the corrections already implemented
in MC generators in order to assess the associated theoretical uncertainty.
\Acknowledgments
We thank Andrea Ferroglia for his invaluable help during all stages of
the computation. We also thank Stefano Pozzorini for discussions and clarifications concerning
the interplay between infrared poles and rational terms, R.~Keith Ellis and Giulia Zanderighi for 
useful communications concerning {\sc QCDLoop} and Andr\'e van Hameren for
help with {\sc OneLOop}. Feynman diagrams have been drawn with the packages 
{\sc Axodraw}~\cite{Vermaseren:1994je} and {\sc Jaxo\-draw}~\cite{Binosi:2003yf}.

The research of S.A. was supported by the Deutsche Forschungsgemeinschaft through 
Sonderforschungsbereich Transregio 9 {\it Computergest\"utzte Theoretische Teilchenphysik},
and in part by the BMBF through the {\it Nutzungsinitiative CERN}. 
The work of G.O. was supported in part by the NSF Grant No.~PHY-0855489.
S.A. and G.O. gratefully acknowledge the hospitality of CERN
at various stages of this work.


\end{document}